\documentclass[aps%
 ,secnumarabic%
,preprint
,amssymb, amsmath,nobibnotes, aps, pre, floatfix,superscriptaddress]{revtex4-1}
\usepackage{graphicx,epsfig}
\usepackage{amsmath}
\usepackage{amsfonts}
\usepackage{fancyhdr}
\usepackage{hyperref}
\usepackage{dirtytalk}
\usepackage{epstopdf}
\usepackage{longtable}
\usepackage{cleveref}
\usepackage{gensymb}
\usepackage[T1]{fontenc}
\usepackage[running]{lineno}

\begin{document}
\pagestyle{fancy}
\title{No Significant Effect of Coulomb Stress on the Gutenberg-Richter Law after
the Landers Earthquake}

%\author{V\'ictor Navas-Portella, \'Alvaro Corral, Abigail Jim\'enez} % Author(s)
%\address{Centre de Recerca Matem\`atica, Universidad de Granada} % Institution(s)

\author{V\'ictor Navas-Portella}
\affiliation{
Centre de Recerca Matem\`atica,
Edifici C, Campus Bellaterra,
E-08193 Barcelona, Spain}
\affiliation{Barcelona Graduate School of Mathematics,
Edifici C, Campus Bellaterra,
E-08193 Barcelona, Spain}
\affiliation{Facultat de Matem\`atiques i Inform\`atica, Universitat de Barcelona, Barcelona, Spain}

\author{Abigail Jim\'enez}
\affiliation{Departamento de Computaci\'on e Inteligencia Artificial. Universidad de Granada. Campus Ceuta
C/. Cortadura del Valle s.n., 51001 Ceuta, Spain}.

\author{\'Alvaro Corral}
\affiliation{
Centre de Recerca Matem\`atica,
Edifici C, Campus Bellaterra,
E-08193 Barcelona, Spain}
\affiliation{Barcelona Graduate School of Mathematics,
Edifici C, Campus Bellaterra,
E-08193 Barcelona, Spain}
\affiliation{Departament de Matem\`atiques,
Universitat Aut\`onoma de Barcelona,
E-08193 Barcelona, Spain}
\affiliation{Complexity Science Hub Vienna,
Josefst\"adter Stra$\beta$e 39,
1080 Vienna,
Austria}

\begin{abstract}
Coulomb-stress theory has been used for years in seismology to understand how earthquakes trigger each other. 
Whenever an earthquake occurs, the stress field changes, and places with positive increases are brought closer to failure.
Earthquake models that relate earthquake rates
and Coulomb stress after a main event, such as the rate-and-state model, assume 
that the magnitude distribution of earthquakes 
%magnitudes 
is not affected 
by the change in the Coulomb stress. By using different slip models, we calculate the change in Coulomb stress in the fault plane for every aftershock after the Landers event (California, USA, 1992, moment magnitude 7.3).
Applying several statistical analyses to test whether the distribution of magnitudes is sensitive to the sign of the Coulomb-stress increase we conclude that no significant effect is observable.
%in particular, the $b-$value of the Gutenberg-Richter law 
%is 
%significantly
%decreases for events 
%that received a decrease in the 
%
Further, whereas the events with a positive increase of the stress are characterized by a much larger proportion of strike-slip events in comparison with the seismicity previous to the mainshock, the events happening despite a decrease in Coulomb stress show no relevant differences in focal-mechanism distribution with respect to previous seismicity.
\end{abstract}

\maketitle

\section*{Introduction}
Since the L'Aquila event in 2009 seismologists have advocated the modeling and testing of earthquakes within a rigorous statistical framework \cite{Jor11}, following on the CSEP (Collaboratory for the Study of Earthquake Predictability) previous works. 
A recent 
%computational??? experiment 
pseudo-prospective forecast
was conducted on the 2010-2012 Canterbury, New Zealand, series, in order to test a total of fourteen earthquake models \cite{Werner2015,MW18}. Its results offer some encouragement for a physical basis in earthquake forecasting and suggest that some of the recent physics-based and hybrid model development have added informative components \cite{Cattania2018}. 

%\begin{figure}
%\includegraphics[width=\linewidth]{RJmodel.jpg}
%\caption{Reasenberg and Jones (1989) model \cite{reasenberg_jones:1989}, from \cite{doi:10.1785/01200809}}
%\label{RJmodel}
%\end{figure}

Our basic understanding of earthquake physics is that stress is being accumulated on certain regions due to different mechanisms, and that those regions rupture whenever that stress surpasses the strength of the material. 
That rupture is the earthquake. 
The mechanisms by which stresses change are diverse:
in addition to tectonic driving,
they can be induced by precedent earthquakes \cite{Stein1992,King1994,Steacy2005,Nandan2016,Ishibe2015}, by volcanic activity \cite{JGRB:JGRB50413}, or even by artificial means, such as injection of fluids \cite{GonzalezCastor2014} or aquifer withdrawal \cite{Shirzaei1416}. 
Coulomb-stress theory has been used to forecast spatial patterns of aftershock rates,  as well as assessing the likelihood of earthquake rupture sequences \cite{Quigley2019,Par00}.
Although there exist instances where its predictive skills are arguable \cite{Hardebeck_jgr, Marsan2003, doi:10.1029/2005JB004076,doi:10.1029/2004JB003277}, the monitoring of the changes in the stress field represents a valuable information for seismic and volcanic hazard forecasting and to proposing the adequate mitigation measures.

A hallmark of statistical seismology and of earthquake hazard assessment 
is the well-known Gutenberg-Richter relation, or Gutenberg-Richter law
\cite{GR44,Utsu_GR,Kagan_book}.
This law states that
earthquake magnitudes must be described in terms of a probability distribution
and that, above a lower cut-off value, 
this distribution is exponential.
In terms of the probability density $f(m)$ one has
\begin{linenomath}
$$
f(m)  = (b \ln 10) 10^{-b(m-m_{min})}
\propto 10^{-b m},
$$
\end{linenomath}
defined for $m\ge m_{min}$
(values below $m_{min}$ are disregarded),
with $m$ the magnitude,
$m_{min}$ the lower cut-off in magnitude,
$b$ the so called $b-$value
(directly related to the exponent $\beta$ of the power-law cumulative distribution of seismic moment,
$\beta=2b/3$),
and the symbol $\propto$ denoting proportionality. (This relation is usually called magnitude-frequency distribution in seismology).
%and ensuring proper normalization.
A straightforward property of the exponential distribution leads to the fact that
the rate (the number per unit time of earthquakes above a certain magnitude $m$)
is also a decreasing exponential function of the magnitude, with the same $b-$value.

Earthquake hazard forecasts usually comprise two stages: 
in the first one, the rate of events is forecasted, while in the second one, the Gutenberg-Richter law is applied to those rates in order to obtain the probabilities of occurrence for each magnitude threshold. 
%%%(Fig. \ref{RJmodel}). ???
In the case of physics-based models, the forecasted rates of events depend on the Coulomb stresses calculated in the region of interest. These models are variants of the rate-and-state model by Dieterich \cite{JGRB:JGRB9328},
\begin{linenomath}
\begin{equation}
{R(t)} = {r} \left[ 1+ \left( e^{-{\Delta CFS}
/B
%{A\sigma} 
}-1 \right)e^{-{t}/{t_{a}}} \right]^{-1}
\label{Eq:rateandstate}
\end{equation}
\end{linenomath}
where 
$R(t)$ is the rate of events (i.e., aftershocks) at any given time $t$ after a mainshock, 
$r$ is the rate of background seismicity, 
$\Delta CFS$ is the increase in Coulomb stress induced by the mainshock, 
%$A$ and $\sigma$ are 
$B$ is a constant, for our purposes, 
and $t_{a}$ is the characteristic relaxation time \cite{JGRB:JGRB9328}.

Note that 
in the application of the Gutenberg-Richter law to the forecasted rate $R(t)$
given by the previous expression it is implicit
that the Coulomb-stress change caused by a mainshock 
does not alter the fufillment of the Gutenberg-Richter law for the aftershocks,
in particular, this law remains the same no matter whether $\Delta CFS$ is positive or negative.
In some sense, 
$R(t)$ inherites the dependence of the background rate $r$ with the magnitude.
Therefore,
the rate-and-state formulation \cite{JGRB:JGRB9328,Par00,Tod05,Cha12,Cat14,Cat15} 
assumes the fulfillment of the Gutenberg-Richter law for the incoming events (aftershocks), 
with no change in the $b-$value.
%predicts that the functional form that relates the changes in rates and changes in stresses is the same, 
%whatever %the stress quantity might be (Eq. \ref{Eq:rateandstate}). 
%the magnitude of the incoming events (i.e., aftershocks).
This assumption is made when inverting earthquake rates to obtain stress changes \cite{SA00,DCO00,JGRB:JGRB50413}. 
Physics-based models also assume the magnitude distribution does not depend on the stress values, so that forecasted rates can be translated into probabilities of occurrence for different magnitudes.

In fact, it has been long debated \cite{Kamer} whether the value of $b$ in the Gutenberg-Richter law
is essentially universal \cite{Kagan_book} or
whether, on the contrary, it is affected by different geophysical conditions.
%such as focal mechanism or stress
%in particular,
Some studies \cite{Sch05,Nar09} have correlated the $b-$value 
%\cite{GR44,Utsu_GR,Kagan_book} 
(and also the parameters of the Omori law \cite{Omo94,Uts61,Utsu_omori}) 
with the style of faulting \cite{Wyss1334}.  
These studies indicate that 
(at least for California, for a long time period)
%$b\approx1.16\pm0.05$ for normal events, 
%$b\approx0.96\pm0.05$ for strike-slip events, 
%and
%$b\approx0.76\pm0.05$ for thrust events 
%$b\approx1.11$ for normal events, 
%$b\approx1.0$ for strike-slip events, 
%and
%$b\approx0.72$ for thrust events 
% 45 0.79 ^ 0.02 0.87 ^ 0.01 1.03 ^ 0.03
$b\approx1.03$ for normal events, 
$b\approx0.87$ for strike-slip events, 
and
$b\approx0.79$ for thrust events 
\cite{Sch05}.
As
the $b-$value is directly related to the log-ratio between the number of small and large earthquakes, 
variations in $b$ can be associated with the ability of an earthquake rupture to propagate 
(more large events, low $b$) or not (less large events, high $b$). 
% once nucleated. 

According to Mohr-Coulomb theory \cite{Nar09,KB04}, 
%for the same coefficient of static friction, 
%the differential shear stress should be much higher for thrust faulting than for normal faulting. 
thrust faults rupture at much higher stress than normal faults
(with strike-slip faults in between, assuming the same value for the coefficient of static friction).
When the stress required to initiate a rupture is higher, 
stress interactions are enhanced and
cracks can propagate faster in many different directions, 
yielding larger earthquakes \cite{Nar09},
%Henceforth, thrust earthquakes have a higher probability to propagate further than normal ones, 
consistent with the empirically observed $b-$values for thrust faulting \cite{Sch05}.
Conversely, for lower rupture thresholds, one should find indeed the large $b-$values 
characterizing normal faulting.
%that the $b$ value may be lower in a compressional regime than in an extensional one. 
%
Although the threshold for triggering might be different for the different styles of faulting, 
the rupture or not of a fault also depends on its previous state. 

Here we investigate, with rigorous statistical tools, 
%if the functional form relating stress and rates depends on the stress values. 
%The first step is to check 
if the Gutenberg-Richter law is affected by the binary choice between positive and negative 
increases of the Coulomb stress, using the sequence of events after the 1992 Landers earthquake.
The next section explains the seismic catalog and the spatio-temporal window used to define this sequence.
Section 3 develops the procedure to calculate the increase in the Coulomb stress
that the Landers earthquake
provokes
in the fault plane of each event in the sequence.
The statistical analysis is also exposed in this section.
Section 4 presents the results and Sec. 5 summarizes the conclusions.
%\comentario{MORE!!!!!}

\section*{Data}

The June 28, 1992, Landers earthquake, 
with a moment magnitude $m=M_w=7.3$ 
and a rake angle $\rho=-177^\circ$,
corresponding to strike-slip focal mechanism,
has been the strongest one in Southern California at least since 1952.
The earthquake
and its subsequent 
aftershock sequence have been extensively studied \cite{Hill1617,JGRB:JGRB9241,Gombberg2001}, 
with a number of slip distributions that describe its rupture \cite{Wald01061994,bruno1999,Steacy2004,Spotila1995}. In this work we use four slip models to calculate the strain; these models are:
Wald and Heaton (referred here to as \textbf{wald})\cite{Wald01061994},
Hernandez et al. (\textbf{hernandez}) \cite{bruno1999},
Landers Big-Bear California (\textbf{lbbcal}) \cite{Steacy2004} and
Landers Surface Rupture (\textbf{lsurfrup}) \cite{Spotila1995}. The terminology is the same as the one used in Ref. \cite{Steacy2004}.

High quality catalogs for Southern California are nowadays available \cite{2012BuSSA.102.2239H,PhysRevE.92.022808};
in particular
in this paper 
we will select the Landers' aftershocks from the 
{Yang}-{Hauksson}-{Shearer} (YHS) catalog 
\cite{2012BuSSA.102.1179Y}, which incorporates focal-mechanism solutions. 
Given the distribution of acceptable mechanisms, the preferred solution is the most probable one  \cite{Hardebeck2002}. The ambiguity of the actual fault plane is solved by considering that the preferred nodal planes
are those associated with the preferred solution listed in the catalog \cite{2012BuSSA.102.1179Y}.
The focal mechanism, in concrete, the rake angle, together with Landers stress field derived from the slip model,
allows us to calculate Coulomb-stress increases (positive or negative) induced by
the mainshock on the actual orientations of the aftershock ruptures. Note that the YHS catalog does not report the moment magnitude necessarily but a preferred magnitude.

In order to better detect the influence of the Landers stress change 
we take a time window of 
100 days after the mainshock
and a spatial window going from 10 to 150 km from the Landers rupture. Landers earthquake is taken as the mainshock for all the slip models except for the lbbcal whose mainshock is the Big-Bear earthquake (which occurred approximately three hours after Landers earthquake with a moment magnitude $m=M_{w}=6.3$ and rake angle $\rho=-180^{\circ}$\cite{Steacy2004}). We tried other choices for the limits of the window finding similar results as reported in the Supplementary Material. This spatio-temporal window defines Landers aftershocks for our purposes. Distances to the fault are computed as the minimum Euclidean distance from the aftershock hypocenter to the center of each fault patch as given by the slip model.
The reason to exclude events closer than 10 km is the uncertainty of the deformation field
near the edges of the subfaults \cite{Okada92}, as the finite-fault approximation provides spurious values near the fault zone because of boundary effects.

\section*{Procedure}

The dMODELS software in Ref.~\cite{Battaglia20131} calculates the deformation field 
(or displacement)
caused by different models corresponding to different physical processes. 
Although there exist many programs that calculate deformation caused by earthquakes, 
this package has been thoroughly tested, and can introduce many different sources of deformation, 
which can be translated into stress changes in a straightforward way. The dMODELS software will be the one used here to obtain deformation field from the different slip models of Landers. 

The local coordinate system for dMODELS is east-north-up, ENU.
After introducing the corresponding slip model (also called source model) 
for the mainshock of interest (Landers in our case \cite{Steacy2004}) into the dMODELS program
we obtain the projections 
in the ENU axes 
of the deformation field $\vec u$ 
caused by the mainshock
at the position of each aftershock 
(and also at its neighborhood, in order to take spatial derivatives). 
We then obtain the strain tensor associated to $\vec u$ 
by calculating the (symmetrized) gradient of the deformation \cite{Lautrup}, 
whose components are
$\varepsilon_{ij}=(\nabla_i u_j + \nabla_j u_i)/2$
(with a spatial step equal to $1$ km). 

Afterwards, we assume an isotropic and elastic material for calculating the stress tensor \cite{Lautrup}, 
or, more precisely, the contribution of the mainshock to the stress tensor,
$s_{ij}=2 \mu \varepsilon_{ij} + \lambda \delta_{ij} \sum_k \varepsilon_{kk}$,
with $\delta_{ij}$ the components of the identity matrix and
with the Lam\'e elastic moduli given by
$\mu=\lambda=3 \times 10^4$ MPa \cite{KB04} (Poisson ratio $\nu=\lambda \left( \lambda + \mu \right)^{-1}/2=0.25$). Moreover, when calculating the stress induced by previous events (mainshocks) on new events (aftershocks)
it is  necessary to orientate it onto the fault \cite{GRL:GRL16929,TOTS17}, 
so that one can actually evaluate if the new events could have been triggered by the induced stress or not.
Given the fault plane and slip vector of an aftershock,
we calculate the change in the normal $\sigma_n$ and shear (or tangential) $\tau$ stresses in that orientation and position, 
as
\begin{linenomath}
\begin{equation}
\Delta \sigma_n = \sum_{ij} n_i s_{ij} n_j
\mbox{ and }
\Delta \tau= \sum_{ij} \ell_i s_{ij} n_j,
\label{ellas}
\end{equation}
\end{linenomath}
with $n_i$ and $\ell_i$ the components of the normal and slip vectors, 
%(both with norm one), 
respectively.
The formulas to obtain the ENU components of these vectors from the information recorded
in the YHS catalog (strike, dip and rake angles \cite{akirichards}) are given in the Methods section.
Note that
in order to be realistic,
the Coulomb-stress changes
have to be calculated onto the planes of the actual faults \cite{GRL:GRL16929}. 
This contrasts with an approach in which 
Coulomb stresses are calculated onto the so-called optimally oriented planes \cite{King1994},
when the only information available is the regional stress. 
However, optimally oriented planes are imaginary planes that might not correspond to the actual geology.

The Mohr-Coulomb failure criterion \cite{vavryvcuk2015earthquake}
states that the shear stress $\tau$ 
on a fault that ruptures must surpass the critical value $\tau_{c}$, 
which is a linear function of the normal stress,
\begin{linenomath}
\begin{equation}
\tau_{c}=C-\mu' \sigma_{n} 
\end{equation}
\end{linenomath}
with $C$ the cohesion and $\mu'$ the effective fault friction coefficient
(including the contribution of the pore pressure 
\cite{JGRB:JGRB12887,King1994}).
Care must be taken with the convention of signs in the normal stress, which is not the same in geophysics than in solid mechanics (our convection takes the negative sign for compression, this is the reason for the negative sign before $\mu '$).
From this failure criterion it is natural to define the Coulomb stress as
$
CFS=\tau+\mu'\sigma_{n},
$
which signals failure by $CFS > C$. In fact, for pre-existing faults one can consider that the cohesion is nearly zero.
The change in Coulomb stress at the aftershock fault plane due to the mainshock
will be
\begin{linenomath}
\begin{equation}\label{eq:coul}
\Delta CFS=\Delta\tau+\mu'\Delta\sigma_{n},
\end{equation}
\end{linenomath}
with $\Delta \tau$ and $\Delta \sigma_{n}$ coming from Eq. (\ref{ellas}).
Thus,
positive increases of the Coulomb stress bring the fault closer to failure, 
whereas negative increases distance it away from failure.
\begin{figure}
  \centering
  \includegraphics[width=\textwidth]{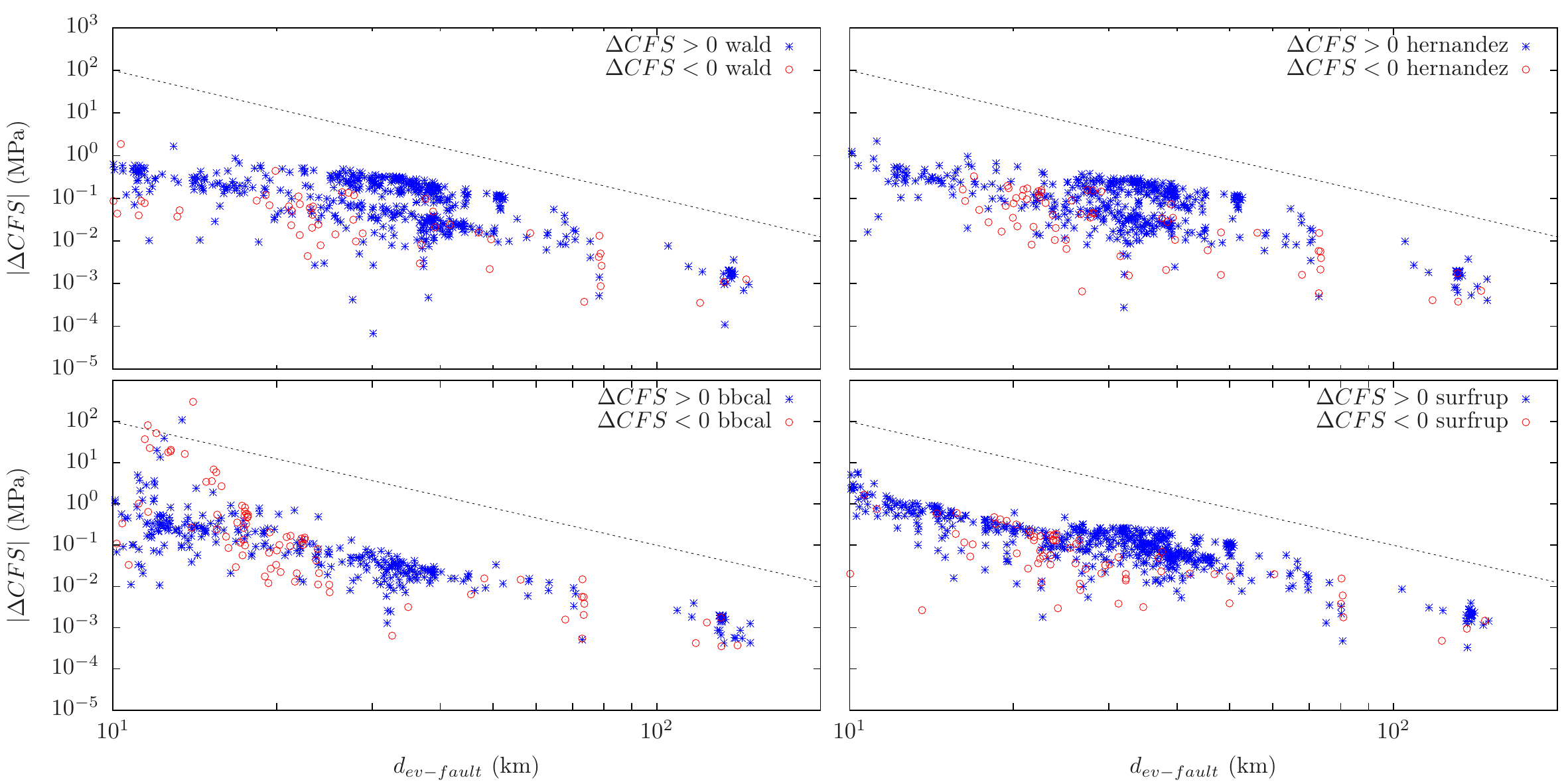}
  \caption{ Dependence of the absolute value of the change in the Coulomb stress $\Delta CFS$ as a function of the distance to the fault of the aftershocks for each slip model, $m_{min}=3$ and $\mu '=0.4$. Aftershocks correspond to the first 100 days after the
Landers mainshock. Distance of aftershocks to the Landers rupture is restricted to be between 10 and 150 km. Black dashed line with slope $-3$, as stated by Coulomb theory, is shown as a guide to the eye.} 
\label{fig:scatter} 
\end{figure}
As the real value of the effective friction coefficient $\mu'$ is uncertain \cite{KB04},
we will check different values of it as in Ref. \cite{Hardebeck_jgr}. In Fig.\ref{fig:scatter} we present a scatter plot that illustrates the dependence of the absolute value of the increase of Coulomb stress $\vert \Delta CFS \vert$ as a function of the distance to the fault $d_{ev-fault}$ for the four slip models. As it is implicit by the Coulomb theory, the value of the increase of Coulomb stress decays as the cube of the distance to the fault. In Fig.\ref{fig:rakes} we show aftershocks with positive and negative increase of Coulomb stress placed in the window we study for the wald slip model and $\mu '=0.4$.

%The complete procedure to obtain the Coulomb stress is summarized in Fig. \ref{fig:flujo}.
%
%\begin{figure}
%  \centering
%  \includegraphics[width=\textwidth]{diagrama_flujo_coulomb_2.pdf}
%  \caption{Flowchart summarizing the procedure to obtain the Coulomb stress on each
%aftershock fault plane from the slip model and the focal-mechanism catalog.
%} 
%\label{fig:flujo} 
%\end{figure}
Once we know the Coulomb-stress change in the fault plane of each aftershock
we can separate these into two subsets attending to the value of the change, with the most natural
separation being between positive and negative increases
(denoted by sub-indexes $>$ and $<$, respectively).
Naturally, we expect to obtain many more aftershocks in positive lobes
than in negative ones \cite{Stein_nature}.
It is 
for each of these subsets that we will study the fulfilment of the Gutenberg-Richter law.
For any set or subset (or sub-catalog) of earthquakes, 
the value of $b$ in the Gutenberg-Richter law 
can be automatically obtained by maximum-likelihood estimation, 
as\cite{Aki_mle,Marzocchi_bvalue}:
%\begin{linenomath}
\begin{equation}
b=\frac {\log_{10} e} {\bar m -m_{min}},
\label{Akiformula}
\end{equation}
%\end{linenomath}
with $\bar m$ the mean magnitude of the events considered
(i.e., those above $m_{min}$). 
Let us stress that $m_{min}$ is not the minimum magnitude recorded in the catalog
but the value from which we fit the Gutenberg-Richter law to the data. As the resolution of the magnitude $\Delta m$ is small ($\Delta m=0.01$) it is not necessary to perform the discreteness correction \cite{Bender1983}.

In principle, results should not significantly depend on the value of $m_{min}$,
but the larger its value the less data to calculate the $b-$value
and the larger the uncertainty,
whereas for a too small $m_{min}$ the Gutenberg-Richter law
would not be fulfilled due to the incompleteness of the catalog 
and the resulting $b-$value would be artefactual. 
In this paper we have taken $m_{min}=3$, 
which ensures the fulfilment of the
Gutenberg-Richter law for all data sets analysed, as we have verified 
by means of the Kolmogorov-Smirnov goodness-of-fit test
\cite{Press}, 
where the distribution of the test statistic 
and, from it, the $p-$value of the fit, $p_{fit}$, 
is calculated using $10^4$ Monte Carlo simulations \cite{Clauset,Corral_Deluca}.
Although some fitting procedures look for the value of $m_{min}$ that optimizes the fit for a given data set \cite{Clauset,Corral_Deluca,CorralGonzalez2019}, 
we have opted for a fixed $m_{min}$ in order to compare the different subsets 
on the same footing.
So, in all cases the exponential fit for $m\ge 3$ cannot be rejected 
($p-$value of the test larger than 0.05).
Note that $m_{min}$ defined in this way can be considered a magnitude of completeness, 
and thus, our value of $m_{min}$ turns out to be rather conservative or strict, 
in the sense that it is larger (and therefore safer) than in other works \cite{Woessner2005}.

The maximum-likelihood estimation of the $b-$value
has an associated uncertainty given by its standard deviation
\begin{linenomath}
$$
\sigma=\frac b {\sqrt{N}},
$$
\end{linenomath}
where $N$ is the number of earthquakes with $m\ge m_{min}$ in the subset,
out of a total number $N_{tot}$ (of any magnitude)\cite{ShiBolt1982}.
Note that this uncertainty only depends on the number of data, 
and has nothing to do with the goodness of the fit.
This result, as well as the formula for the maximum-likelihood estimation of $b$, Eq. (\ref{Akiformula}), can also be obtained from Ref. \cite{Corral_Deluca} just taking into account the relation
between moment magnitude and seismic moment. 
This standard deviation, $\sigma$, is what represents the uncertainty when we report our resulting $b$-values.

The comparison between 
the $b-$values of the subsets with 
different values of 
$\Delta CFS$ is done by means of the following statistic
\begin{linenomath}
$$
z=\frac{b_> - b_<}{\sqrt{\sigma_>^2+\sigma_<^2}}
=\frac{b_> - b_<}{\sqrt{b_>^2/N_> + b_<^2/N_<}},
$$
\end{linenomath}
where the sub-indexes $>$ and $<$ refer to positive and negative increases
of the Coulomb stress.
This statistics is rooted on the null hypothesis that 
both subsets of data (positive and negative) % increases of Coulomb stress)
belong to the same underlying population of earthquake magnitudes
and then, both estimators of the $b-$value ($b_>$ and $b_<$)
have a common mean value, 
which is that of the whole population.
Therefore, under the null hypothesis, $b_>-b_<$ has zero mean
and standard deviation $\sqrt{\sigma_>^2+\sigma_<^2}$
(approximating the population variance from the sample values of $b_>$ and $b_<$
and
assuming zero covariance between $b_>$ and $b_<$)
and then $z$ has zero mean too and unit standard deviation. An additional assumption is that $z$ is normally distributed, 
which is supported by theory in the asymptotic limit 
($N_>$ and $N_<$ going to infinity \cite{pawitan2001}). 
Assuming normality
we will test the null hypothesis just comparing the value of
$z$ with the standard normal distribution
and the hypothesis will be rejected if the value of $z$ is too extreme
for a given significance level;
in quantitative terms
this will be given by a $p-$value, called $p_{norm}$,
smaller than the significance level 
(0.05, let us say; corresponding to 0.95 confidence).

If we 
do not want to believe
that the asymptotic regime has been reached 
the best option is to use a permutation test \cite{Good_resampling}. 
Under the null hypothesis 
(all values of magnitude belong to the same population) 
one is allowed to aggregate both subsets (positive and negative) 
and take, without repetition, two sub-samples of size $N_>$ and $N_<$; 
note that this is equivalent to take a permutation of the aggregated sample
and separate it into two parts ($>$ and $<$).
One proceeds in the same way as in the original data, 
calculating (by maximum likelihood) 
$b_>^*$,  $b_<^*$,
and from here
$\sigma_>^*$, $\sigma_<^*$, and $z^*$,
where the asterisk marks that we are dealing with a permutation 
of the original data.
Repeating the permutation procedure many times we find the distribution of
$z^*$, which can be compared with the original value $z$.
The $p-$value of the permutation test, $p_{perm}$, will be given by the fraction of
permutations for which $|z^*|$ is larger than $|z|$ (the empirical value).
In our case we take $10^4$ permutations.

As a complement, 
instead of the fitted $b-$values
we may directly compare the distributions;
this can be done with the two-sample Kolmogorov-Smirnov test, 
whose null hypothesis is that both data sets come from the same population, 
so, the two empirical distributions ($>$ and $<$) are two realizations
of a unique theoretical distribution (which remains unveiled)  \cite{Press}. This test leads to a $p$-value that we call $p_{2ks}$.
A final comparison comes from the application of the Akaike information criterion ($AIC$) \cite{AICBOOK}.
We consider that we aggregate both subsets (positive and negative $\Delta CFS$)
but keeping the distinction in the sign of $\Delta CFS$.
Then, we contemplate two options.
Model 1, simple:
we fit the aggregated data set with one single Gutenberg-Richter exponential
leading to the value $b_{all}$.
Model 2, ``complex'':
we fit each data set with its own exponential function
(values $b_>$ and $b_<$ in the same table).
In each case, $AIC=2k-2\hat \ell$, where $k$ is the number of parameters
of each model and $\hat \ell$ is the log-likelihood of the model at maximum. The likelihood in model 2 is the sum of likelihoods for each subcatalog \cite{pawitan2001}.
The model yielding the smallest $AIC$ should be prefered. Defining $\Delta AIC= AIC_{2}-AIC_{1}$ leads to the rejection of the simple model when $\Delta AIC$ is significantly below zero (see next section).
%Computing the difference,
%We arrive at 
%$\Delta AIC= AIC_2-AIC_1=2-2(-22867.42+22869.27)=-1.7$;
%so, the ``complex'' model 2, with two separate 
%$b-$values, is preferred in front the simplicity of just one $b-$value
%(on the other hand, a likelihood-ratio test does not find enough evidence in favor of the 
%complex model; this is not contradictory as this test is more strict than the $AIC$).

\section*{Results}

Table \ref{tab:mlefitting} shows the values of $b$ obtained from the application of the maximum likelihood estimation and goodness-of-fit test explained above to the different subcatalogs obtained from the Landers sequence.
We can see how, in the overall case
(when events are not separated in terms of Coulomb-stress change), 
the Gutenberg-Richter law is fulfilled with an average value $b_{all}=0.92$. Each slip model leads to a different value of $b_{all}$ because the fault geometry is different, and events too close to the fault are discarded.
This $b-$value for the Landers aftershocks is found, 
not surprisingly, to be close to the average for aftershocks in California, $b\simeq 0.9$ \cite{Reasenberg_Jones89,Jones1994},
and somewhat below the long-term value of Southern California (all events), $b\simeq 1.0$
\cite{Hutton_Woessner}
(although other works report $b\simeq 1.0$ for Landers aftershocks, 
probably due to the consideration there of a much smaller magnitude of completeness \cite{Shcherbakov2005}).

After separating by the sign of the Coulomb-stress change, 
the first result that becomes apparent from the table
is that the number of aftershocks with positive increases
is much larger than the number for the negative case \cite{King1994,Steacy2005}, 
no matter neither the slip model (nor the value of $\mu'$) used to calculate $\Delta CFS$. 
Regarding the $b-$values, although they depend on the slip model, 
we can summarize them by taking the mean of the four models and taking $\mu'=0.4$ as 
$b_>\simeq 0.93$ and
$b_< \simeq 0.87$ with individual uncertainties around $0.04$ and $0.11$ respectively.
Note that the magnitude distribution for the overall case 
is a mixture of the distributions corresponding to $\Delta CFS > 0$
and $\Delta CFS <0$, and 
therefore, the value of $b$ in the overall case turns out to be the harmonic mean
of $b_>$ and $b_<$, i.e., 
\begin{linenomath}
\begin{equation}
b_{all}^{-1}=\frac{N_> b_>^{-1} + N_< b_<^{-1}}{N_>+N_<},
\label{harmonic}
\end{equation}
\end{linenomath}
see Refs.~\cite{hernandez2014,Navas_pre2}.
%If instead %of the separation between positive and negative 
%of $\Delta CFS > 0$
%we consider $\Delta CFS > 0.1$ MPa, the resulting values of $b$
%are nearly the same, but with a larger uncertainty
%(with little dependence on the value of $\mu'$).
%
Despite the fact the values of $b_>$ and $b_<$ do not look much different between them, 
statistical testing becomes necessary in order to establish significance \cite{Corral_Boleda}. 

\begin{figure}[h!]
\includegraphics[width=\textwidth]{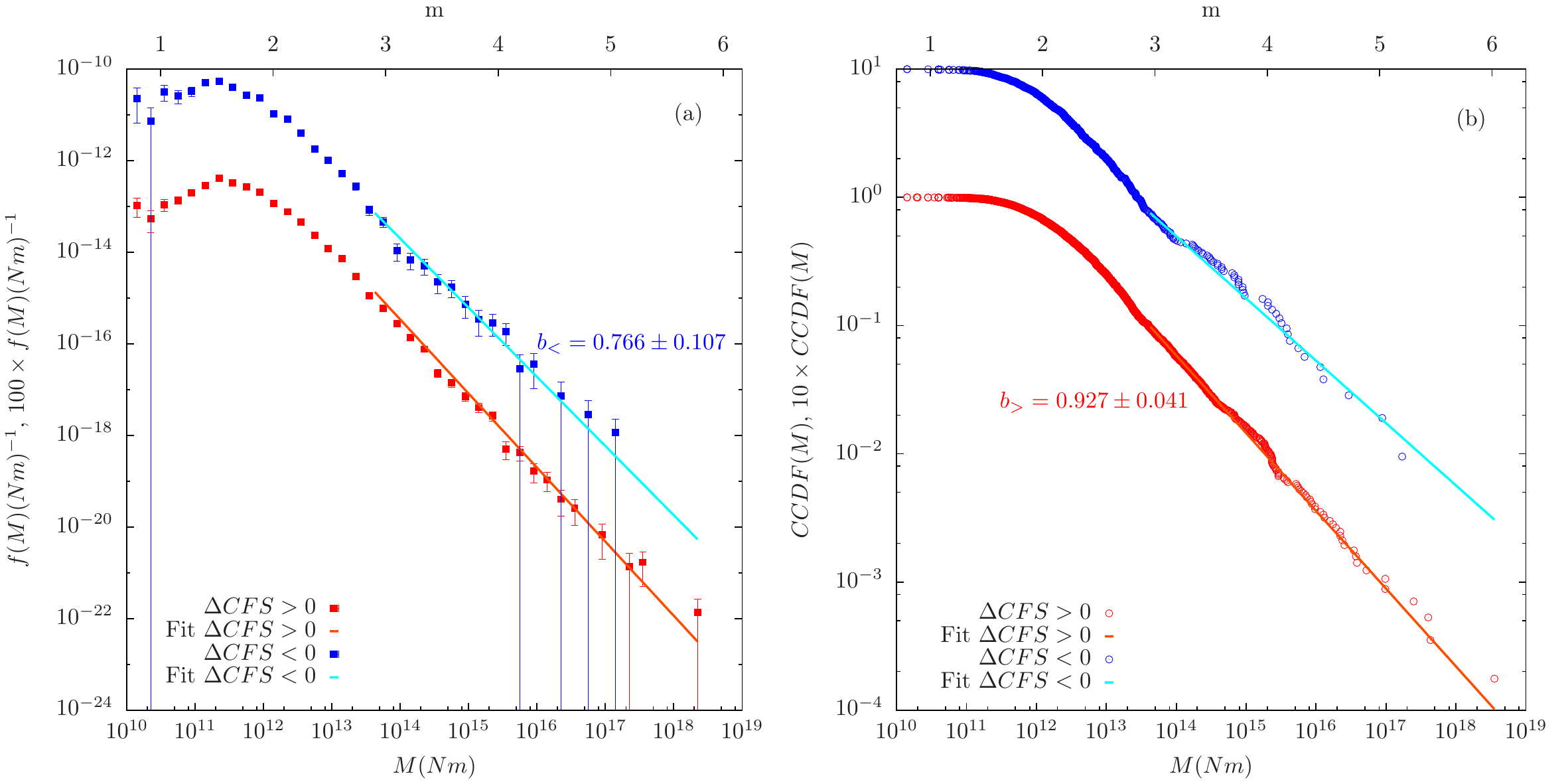}
\caption{\label{fig:distributions} 
  Estimation of the probability densities (a) and of the complementary cumulative distribution functions (CCDF) (b)
of seismic moment $M$
for Landers aftershocks with $\Delta CFS>0$ and $\Delta CFS<0$ for the wald slip model using
$\mu'=0.4$ in the calculation of $\Delta CFS$. 
Curves corresponding to $\Delta CFS<0$ have been conveniently multiplied by a factor 100 and 10 respectively 
for clarity sake.
Error bars in (a) denote one standard deviation, 
and are symmetric, despite the appearance in log scale, 
see Ref. \cite{Corral_Deluca}.}
\end{figure}

Table \ref{tab:permutational} compares $b_>$ and $b_<$ for the different slip models taking $\mu '=0.4$,
and shows that 
the difference in the $b-$values can not be considered 
significantly different from zero with a confidence larger than $0.95$ 
so, the null hypothesis $b_> \simeq b_<$ can not be rejected. This result is true for all the statistical tests as all the $p$-values are greater than $0.05$. Table \ref{tab:permutational} also shows the results of the two-sample Kolmogorov-Smirnov test and the calculation of $\Delta AIC$ leading in both cases to the result that no change in the distributions as a function of positive and negative $\Delta CFS$ can be established. In concrete, $\Delta AIC$  is always greater than the critical value $\Delta AIC_{c}=-1.84$ \cite{AICBOOK,Murtaugh2014} at significance level of $0.05$. 
The wald slip model is the one for which both distributions (positive and negative) appear as more different; however, the difference is not significant. 
Figure \ref{fig:distributions} shows the probability density functions as well as the complementary cumulative probability functions in this case. % Nevertheless, neither in this case the difference between the $b$-values is significant.
%For the comparison between the cases $\Delta CFS > 0.1$ MPa and $\Delta CFS < 0$,
%the fact that the uncertainty of $b_{>0.1}$ is larger than that of $b_>$ 
%(despite the values themselves are rather similar) also leads to
%a non-significant difference between $b_{>0.1}$ and $b_<$, 
%and we cannot reject that $b_{>0.1} \simeq b_<$.

\begin{figure}
  \centering
  \includegraphics[width=\textwidth]{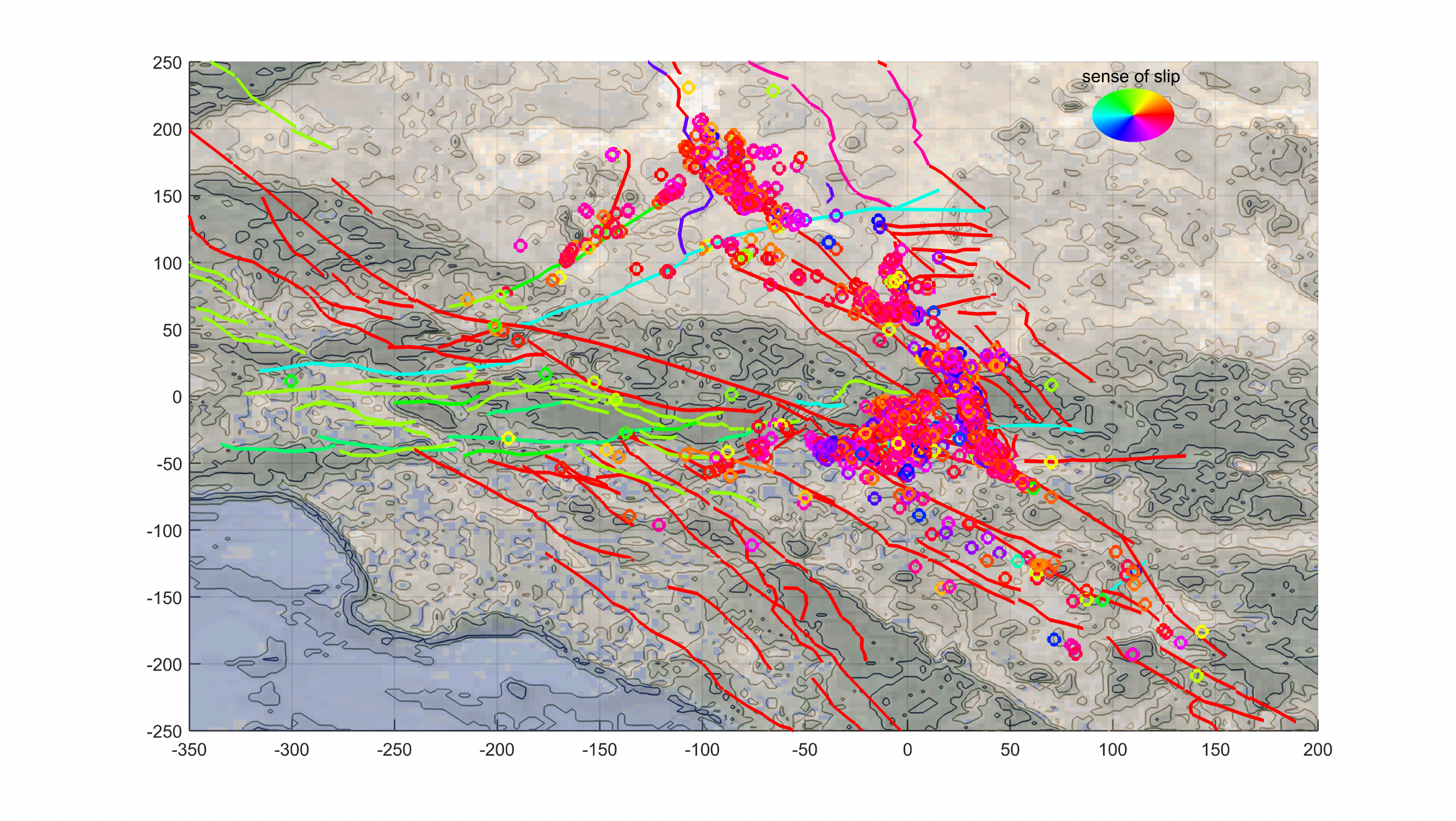}
	\includegraphics[width=\textwidth]{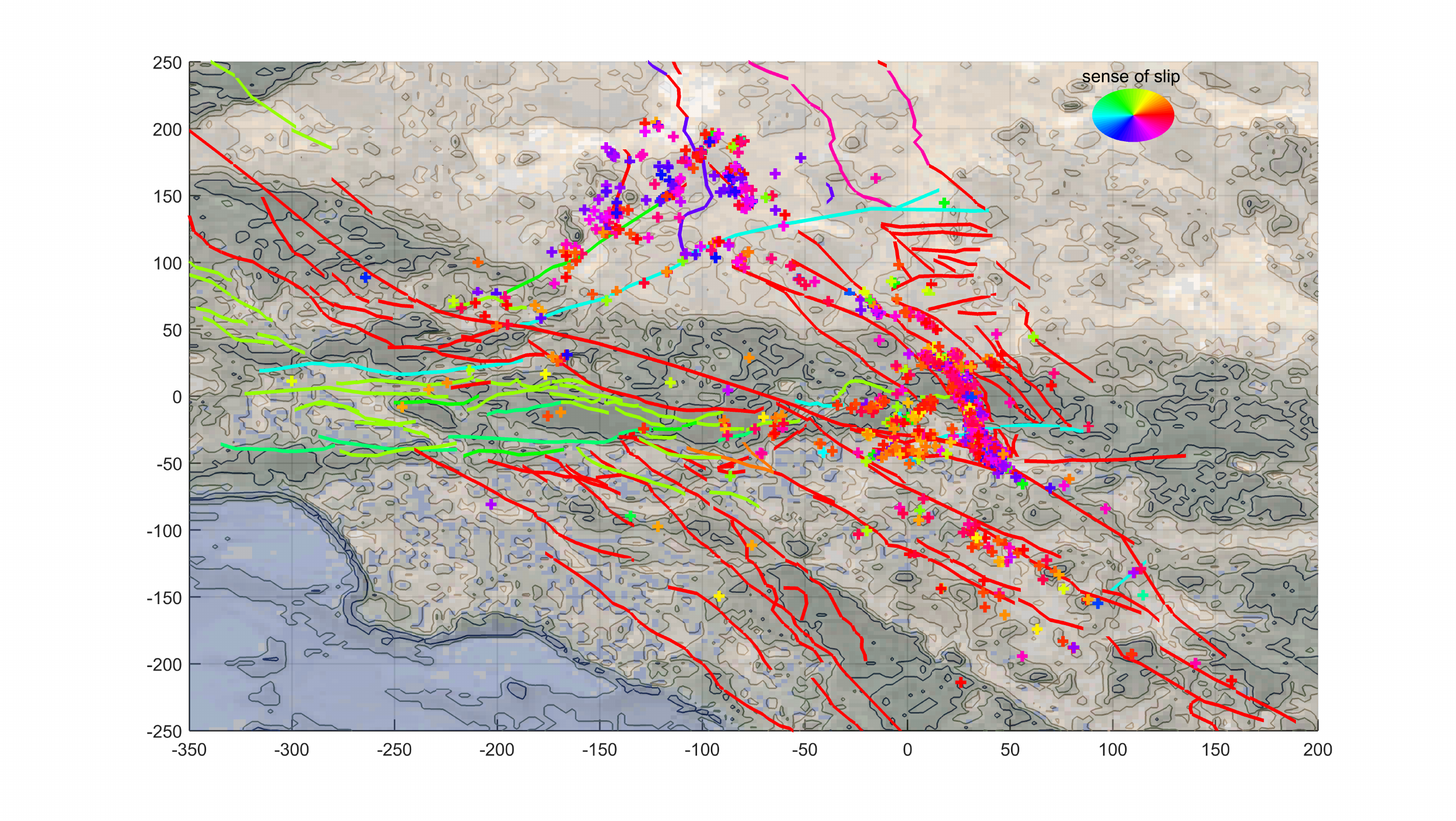}
  \caption{Rakes of Landers aftershocks compared to the fault network
(time window of 100 days after the mainshock).
Top: $\Delta CFS >0$.
Bottom: $\Delta CFS <0$.
Color scale for the sense of slip: 
red for right-lateral ($\rho$ close to $\pm 180^\circ$), 
light blue for left-lateral ($\rho$ close to $0^\circ$), 
green for normal ($\rho$ close to $-90^\circ$) 
and dark blue or purple for thrust faulting
($\rho$ close to $90^\circ$).
No restriction on the magnitude values is used.
An area of $550 \times 500$ km is shown.
%Origin?? km???
} 
\label{fig:rakes} 
\end{figure}

As mentioned in the introduction, some authors have unveiled a direct
dependence of the $b-$value on the focal mechanism of the events,
which implies a dependence of $b$ on the total stress (not the stress increase) \cite{Sch05}.
The rake angle is associated to the focal mechanism in the following way:
values of the rake around $-90^\circ$ correspond to normal events (labelled as $no$),
values around $0^\circ$ or $\pm 180^\circ$ to strike-slip events ($ss$),
and
values around $90^\circ$ to thrust events ($th$). 
We do not find any significant effect of the rake on the $b-$value (See Table \ref{table_fm}),
due to the low number of events in the normal and thrust regimes
(which increases the uncertainty).
But despite the large uncertainty, the values of $b_{no}$ and $b_{ss}$
are roughly in agreement with the results of Ref. \cite{Sch05};
however, our value of $b_{th}$ turns out to be rather large in comparison
(but compatible, within the error bars).
We further observe that ratios $N_{> ss}/N_{< ss}$ and $N_{> no}/N_{< no}$ are higher than $N_{> th}/N_{< th}$; 
i.e., in strike-slip and normal events the contribution from $\Delta CFS > 0$ is higher than in thrust events, 
as can be verified looking at Table \ref{table_fm}.
Comparing with the number of earthquakes with each focal mechanism 
for the 5 years previous to Landers 
%($N_{no}=77$, $N_{th}=61$, and $N_{ss}=498$ for $m>3$)
we conclude that it is indeed the low number of thrust aftershocks with positive $\Delta CFS$
which is anomalous (and not the relatively high number of them for negative $\Delta CFS$),
due to an increase in the number of normal events
and an even higher increase in strike-slip events
triggered ($\Delta CFS>0$) by the Landers mainshock.
This difference in numbers becomes visually apparent in
Fig. \ref{fig:rakes}.
%Although the two populations ($\Delta CFS>0$ and $<0$) are different in terms of focal mechanism, 
%there is no substantial difference in the fulfilling of the Omori law.
%Indeed, if we compare this for the two subsets we find the ``characteristic'' power-law Omori decay 
%of the rate with very similar values of 
%the Omori exponent.
%Note that this is in disagreement with the rate-and-state formulation \cite{JGRB:JGRB9328}, 
%which does not predict Omori behavior in the case of negative $\Delta CFS$.

\begin{table}[htbp]
\begin{tabular}{|l|c|r|r|r|r|r|}
\hline

Slip Model &  \textbf{} & \multicolumn{1}{l|}{\textbf{$N_{tot}$}} & \multicolumn{1}{l|}{\textbf{${N}$}} & \multicolumn{1}{l|}{\textbf{$b-$value}} & \multicolumn{1}{l|}{\textbf{$\sigma$}} & \multicolumn{1}{l|}{\textbf{$p_{fit}$}} \\ \hline \hline
%  Overall & 6730 & 662 & $b_{all}=0.882$ & 0.034 & $0.233 \pm 0.004$ \\ \hline \hline
%%%%%%%%%%%%%%%
\textbf{wald} &  \textbf{$\Delta CFS>0$} & 5213 & 509 & $b_>=0.927$ & 0.041 & $0.313 \pm 0.005$ \\ %\hline
\phantom{\textbf{wald}}  & \textbf{$\Delta CFS<0$} & 814 & 51 & $b_<=0.766$ & 0.107 & $0.861 \pm 0.003$ \\ 
\phantom{\textbf{wald}}  & All & 6027 & 560 & $b_{all}=0.909$ & 0.038 & $0.243 \pm 0.004$ \\ \hline
\textbf{hernandez} &  \textbf{$\Delta CFS>0$} & 5027 & 465 & $b_>=0.926$ & 0.043 & $0.505 \pm 0.005$ \\ %\hline
\phantom{\textbf{hernandez}} & \textbf{$\Delta CFS<0$} & 765 & 62 & $b_<=0.866$ & 0.110 & $0.197 \pm 0.004$ \\ 
\phantom{\textbf{hernandez}}  & All & 5792 & 527 & $b_{all}=0.919$ & 0.040 & $0.231 \pm 0.004$ \\ \hline
\textbf{bbcal} & \textbf{$\Delta CFS>0$} & 3641 & 309 & $b_>=0.978$ & 0.056 & $0.232 \pm 0.004$ \\ %\hline
\phantom{\textbf{bbcal}} & \textbf{$\Delta CFS<0$} & 1191 & 82 & $b_<=0.948$ & 0.105 & $0.327 \pm 0.005$ \\
\phantom{\textbf{bbcal}}  & All & 4832 & 391 & $b_{all}=0.971$ & 0.049 & $0.053 \pm 0.002$ \\ \hline
\textbf{surfrup} &  \textbf{$\Delta CFS>0$} & 5534 & 548 & $b_>=0.890$ & 0.038 & $0.290 \pm 0.005$ \\ %\hline
\phantom{\textbf{surfrup}} & \textbf{$\Delta CFS<0$} & 774 & 68 & $b_<=0.891$ & 0.108 & $0.555 \pm 0.005$ \\ 
\phantom{\textbf{surfrup}}  & All & 6308 & 616 & $b_{all}=0.890$ & 0.036 & $0.239 \pm 0.004$ \\ \hline

\end{tabular}
\caption{
Results of fitting the 
Gutenberg-Richter law to the Landers aftershocks, separating positive and negative Coulomb-stress increases,
for different Slip models, $\mu '=0.4$
and $m_{min}=3$.
Aftershocks correspond to the first 100 days after the Landers mainshock.
Distance of aftershocks to the Landers rupture is restricted to be between 10 and 150 km.
The $p-$value of the goodness-of-fit test is computed with $10^4$ simulations
and is denoted by $p_{fit}$. 
Its uncertainty corresponds to one standard deviation. In no case the Gutenberg-Ricther law can be rejected.
}
\label{tab:mlefitting}
\end{table}

%
%\begin{figure}[h!]
%\center
%\includegraphics[scale=0.75]{distributionsGRdis5.eps}
%\includegraphics[scale=0.75]{cdfs.eps}
%\caption{\label{fig:distributions} 
%Estimation of the probability densities (a)
%and of the 
%complementary cumulative distribution functions (CCDF) (b) and (c)
%of seismic moment $M$
%for Landers aftershocks with $\Delta CFS>0$ and $\Delta CFS<0$, using
%$\mu'=0.4$ in the calculation of $\Delta CFS$. (b) and (c) contain essentially the same information, displayed in different ways.
%Curves corresponding to $\Delta CFS<0$ have been conveniently multiplied by a factor in (a) and (b) 
%for clarity sake. A vertical line in (c) denotes where the Kolmogorov-Smirnov statistic is attained.
%}
%\end{figure}

\begin{table}[htbp]
\begin{tabular}{|c||c|r|r|| r|r||r||}
\hline
Slip Model &
$z$ & \multicolumn{1}{l|}{\textbf{$p_{norm}$}} & \multicolumn{1}{l||}{\textbf{$p_{perm}$}} 
& $d_{2ks}$ & $p_{2ks}$ & $\Delta AIC$  \\ \hline\hline
\textbf{wald} & 1.396 & 0.163 & $0.156 \pm 0.004$  & 0.139 & 0.311 & 0.234  \\ %\hline
\textbf{hernandez} & 0.511 & 0.609 & $0.596 \pm 0.005$ & 0.095 & 0.690 & 1.748 \\ %\hline
\textbf{bbcal} & 0.254 & 0.800 & $0.828 \pm 0.004$ & 0.063 & 0.952 & 1.936 \\ %\hline
\textbf{surfrup} & -0.010 & 0.992 & $0.994 \pm 0.001$ & 0.094 & 0.643& 1.999 \\ \hline
\end{tabular}
\caption{
Results of the statistical tests comparing $b$-values and
magnitude distributions for positive and negative Coulomb-stress changes,
using different slip models and $\mu '=0.4$. Values of $\Delta AIC= AIC_{2}-AIC_{1}$ are also included. Same data as previous table.
Columns 2 to 4: testing the null hypothesis that there is no difference between the $b-$values 
(i.e., $b_>=b_<$).
Columns 5 to 6: testing the null hypothesis that there is no difference in the distributions, 
using the 2-sample Kolmogorov-Smirnov test.
In the first test, both asymptotic normality of the $z$ statistic and a permutation test are used for the calculation of the $p-$value
(labeled as $p_{norm}$ and $p_{perm}$, respectively).
In the latter case the number of permutations is $10^4$, 
and 
the uncertainty of $p_{perm}$ corresponds to one standard deviation. 
$d_{2ks}$ and $p_{2ks}$ are the Kolmogorov-Smirnov statistic and its $p-$value.}
\label{tab:permutational}
\end{table}

\begin{table}[htbp]
\begin{tabular}{|l||r|r|r|r||r|r|r|r||}
\hline
fm & \multicolumn{1}{l|}{\textbf{$N_{> fm}$}} & \multicolumn{1}{l|}{\textbf{$N_{< fm}$}} & \multicolumn{1}{l|}{\textbf{$N_{fm}$}} & 
\multicolumn{1}{l||}{$N_{fm}^{pre}$}& \multicolumn{1}{l|}{\textbf{$b_{> fm}$}} & \multicolumn{1}{l|}{\textbf{$b_{< fm}$}} & \multicolumn{1}{l|}{ $b_{fm}$} & \multicolumn{1}{l||}{ $b_{fm}^{pre}$} \\ \hline \hline
\textbf{wald} & \multicolumn{1}{l|}{} & \multicolumn{1}{l|}{} & \multicolumn{1}{l|}{} & 
\multicolumn{1}{l||}{} & \multicolumn{1}{l|}{} & \multicolumn{1}{l|}{}& \multicolumn{1}{l|}{}& \multicolumn{1}{l||}{} \\ \hline \hline
No: $-135^\circ \le \rho \le -45^\circ$
& 39 & 5 & $N_{no}=\phantom{0}44$ & $N^{pre}_{no}=\phantom{0}77$ & 1.047 & - & $b_{no}=1.081$ & $b_{no}^{pre}=1.484$ \\ 
Th: $\phantom{-1}45^\circ \le \rho \le 135^\circ$ %\textbf{Thrust} 
& 9 & 8 & $N_{th}=\phantom{0}17$ & $N^{pre}_{th}=\phantom{0}77$ & - & - & $b_{th}=0.929$  & $b_{th}^{pre}=0.899$ \\  
SS: the rest
& 461 & 38 & $N_{ss}=499$ & $N^{pre}_{ss}=\phantom{0}503$ & 0.914 & 0.726 & $b_{ss}=0.896$  & $b_{ss}^{pre}=0.960$ \\ \hline  \hline
\textbf{hernandez} & \multicolumn{1}{l|}{} & \multicolumn{1}{l|}{} & \multicolumn{1}{l|}{} & 
\multicolumn{1}{l||}{} & \multicolumn{1}{l|}{} & \multicolumn{1}{l|}{}& \multicolumn{1}{l|}{}& \multicolumn{1}{l||}{} \\ \hline \hline
No: $-135^\circ \le \rho \le -45^\circ$
& 38 & 3 & $N_{no}=\phantom{0}41$ & $N^{pre}_{no}=\phantom{0}78$ & 0.995 & - & $b_{no}=1.011$  & $b_{no}^{pre}=1.502$ \\ 
Th: $\phantom{-1}45^\circ \le \rho \le 135^\circ$ %\textbf{Thrust} 
& 7 & 10 & $N_{th}=\phantom{0}17$ & $N^{pre}_{th}=\phantom{0}61$ & - & - & $b_{th}=0.929$  & $b_{th}^{pre}=0.899$ \\  
SS: the rest
& 420 & 49 & $N_{ss}=469$ &  $N^{pre}_{ss}=\phantom{0}506$ & 0.920 & 0.840 & $b_{ss}=0.911$  & $b_{ss}^{pre}=0.957$ \\ \hline  \hline
\textbf{bbcal} & \multicolumn{1}{l|}{} & \multicolumn{1}{l|}{} & \multicolumn{1}{l|}{} & 
\multicolumn{1}{l||}{} & \multicolumn{1}{l|}{} & \multicolumn{1}{l|}{}& \multicolumn{1}{l|}{}& \multicolumn{1}{l||}{} \\ \hline \hline
%\textbf{Normal} 
No: $-135^\circ \le \rho \le -45^\circ$
& 22 & 4 & $N_{no}=\phantom{0}26$ & $N^{pre}_{no}=\phantom{0}80$ & 1.128 & - & $b_{no}=1.165$  & $b_{no}^{pre}=1.521$ \\ 
Th: $\phantom{-1}45^\circ \le \rho \le 135^\circ$ %\textbf{Thrust} 
& 7 & 5 & $N_{th}=\phantom{0}12$ & $N^{pre}_{th}=\phantom{0}69$ & - & - & $b_{th}=1.309$  & $b_{th}^{pre}=0.831$ \\  
SS: the rest
& 280 & 73 & $N_{ss}=353$ & $N^{pre}_{ss}=\phantom{0}507$ & 0.970 & 0.886 & $b_{ss}=0.951$  & $b_{ss}^{pre}=0.958$ \\ \hline  \hline
\textbf{surfrup} & \multicolumn{1}{l|}{} & \multicolumn{1}{l|}{} & \multicolumn{1}{l|}{} & 
\multicolumn{1}{l||}{} & \multicolumn{1}{l|}{} & \multicolumn{1}{l|}{}& \multicolumn{1}{l|}{}& \multicolumn{1}{l||}{} \\ \hline \hline
No: $-135^\circ \le \rho \le -45^\circ$
& 46 & 5 & $N_{no}=\phantom{0}51$ & $N^{pre}_{no}=\phantom{0}76$ & 0.939 & - & $b_{no}=0.966$  & $b_{no}^{pre}=1.470$ \\ 
Th: $\phantom{-1}45^\circ \le \rho \le 135^\circ$ %\textbf{Thrust} 
& 9 & 11 & $N_{th}=\phantom{0}20$ & $N^{pre}_{th}=\phantom{0}61$ & - & 1.010 & $b_{th}=0.886$  & $b_{th}^{pre}=0.899$ \\  
SS: the rest
& 493 & 52 & $N_{ss}=545$ & $N^{pre}_{ss}=\phantom{0}504$ & 0.888 & 0.844 & $b_{ss}=0.884$  & $b_{ss}^{pre}=0.961$ \\ \hline  \hline
\end{tabular}
\caption{
Number of events and $b-$values corresponding to Landers aftershocks with $m \geq 3$
separated by
sign of the Coulomb-stress increase ($>$ and $<$)
and by focal mechanism ($fm$) for each slip model.
$fm=no$ (normal), $ss$ (strike-slip), and $th$ (thrust).
The Coulomb stress is calculated with $\mu'=0.4$.
Same data as in previous tables. Values of $b$ calculated with 10 or less events are not reported. Values for the 5 years previous to Landers are also included. 
}
\label{table_fm}
\end{table}

\section*{Discussion}
We have seen how 
the positive Coulomb-stress increase associated to
the Landers mainshock
triggered a very large number of strike-slip events
and also a large number of normal events, 
but much less thrust events.
Although this result seems easy to establish, as 
it can be obtained without the calculation of $\Delta CFS$ 
(due to the fact that most of the events have $\Delta CFS >0$
and thus, this subset dominates the overall statistics),
we have unambiguosly associated these events to the positive $\Delta CFS$. 
On the other side,
the events 
in the opposite regime
(with $\Delta CFS< 0$) keep a proportion between
normal, strike-slip, and thrust events 
rather different to the $\Delta CFS >0$ case, 
and close to that of the immediately previous record 
(1987-1992, up to Landers).
These results are largely independent on the slip model 
%as well as the value of $\mu'$ 
used to calculate the change in Coulomb stress.
%
%AQUI FALTA UN PARRAFO RESUMIENDO LOS RESULTADOS ESTADISTICOS!!
We have also found that
the $b$-values of the Gutenberg-Richter law
for events with positive $\Delta CFS$ (for which $b_> \simeq 0.93$) 
are in general larger than 
the $b-$values for the events with negative $\Delta CFS$ 
($b_< \simeq 0.87$);
nevertheless, this difference is not statistically significant for any of the slip models used to compute the change in the Coulomb stress.

A number of extensions and improvements could be incorporated 
to our approach in future research. Moreover, we need to take into account the relation between $b$-values and differential stress \cite{amitrano2003}.
We make use of slip models with relatively low resolution in space;
so, it would be interesting to know if higher resolution slip models \cite{Olsen1997, Peyrat2001}
lead to somewhat different values of the strain and the stress, 
in particular close to the fault.
%This could influence the $b-$values.
Also, some authors have argued that 
real faults should have rather low values
of the $\mu'$ coefficient \cite{Mulargia2016}. 
We provide some check of this in the supplementary material, 
which leads to the
conclusion that $\mu'$ has little influence on the $b-$values.
Further, in our temporal window of 100 days, the effect
of viscoelastic relaxation \cite{Sabadini2016}
should be important;
so, this would need to be incorporated into the calculation of the stress.
Finally, in a preliminary analysis we have seen that
there is no substantial difference in the fulfilling of the Omori law 
\cite{Omo94,Uts61,Utsu_omori} 
in the two populations of events ($\Delta CFS>0$ and $<0$).
Indeed, if we compare this for the two subsets we find the ``characteristic'' power-law Omori decay 
of the rate with very similar values of 
the Omori exponent.
Note that this is in disagreement with the rate-and-state formulation \cite{JGRB:JGRB9328}, 
which does not predict Omori behavior in the case of negative $\Delta CFS$.
Certainly, more research using other mainshocks (for which detailed slip models were available) is necesary, 
in order to reduce the statistical uncertainty by means of aggregated distributions, 
which could lead to the detection of small significant differences in both populations
of events.
%

%Regarding the implications for
%physics-based models of aftershock sequences,
%these models assume the $b-$value in the Gutenberg-Richter law is constant for the whole aftershock zone. This is the foundation for both inverting stress from seismicity rates \cite{SA00,DCO00,JGRB:JGRB50413} and for forecasting the seismicity afterwards 
%(Ref. \cite{Cattania2018} and references therein). 
%However, our results demand for more complete models of aftershock occurrence
%in order to get better forecasts.
%Not only 
%different styles of faulting for the aftershocks should be 
%acknowledged, 
%but also 
%the sign of the Coulomb-stress increase should lead to different parameterizations in the rate-and-state model. 

\section*{Methods}
The YHS catalog characterizes fault planes and slip vectors by means of 
three angles:
strike $\Theta$, dip $\delta$, and rake $\rho$.
%, see Fig. \ref{fig:fault}.
In term of these, 
the normal vector of the fault is given by
\begin{linenomath}
\begin{equation}\label{Eq.normdir}
\hat{n}=\left( \begin{array}{c}
n_{E}\\
n_{N}\\
n_{U}\end{array}\right)= \left( \begin{array}{c}
\cos \Theta \sin \delta \\
-\sin \Theta \sin \delta\\
 \cos \delta \end{array} \right)
\end{equation}
\end{linenomath}
in the ENU coordinate system \cite{Smith2006}.
In the same way,
the slip vector is obtained as
\begin{linenomath}
\begin{equation}\label{Eq.slipdir}
\hat{\ell}=\left( \begin{array}{c}
\ell_{E}\\
\ell_{N}\\
\ell_{U}\end{array}\right)= \left( \begin{array}{c}
\sin \Theta \cos \rho - \cos \Theta \cos \delta \sin \rho \\
\cos \Theta \cos \rho + \sin \Theta \cos \delta \sin \rho  \\
 \sin \delta \sin \rho \end{array} \right).
\end{equation}
\end{linenomath}
Note that $\hat n$ and $\hat \ell$ are unit vectors.

%\bibliographystyle{unsrt}
%\bibliography{sample}

\section*{Supplementary Material}

\begin{figure}[htpb]
\includegraphics[scale=0.75]{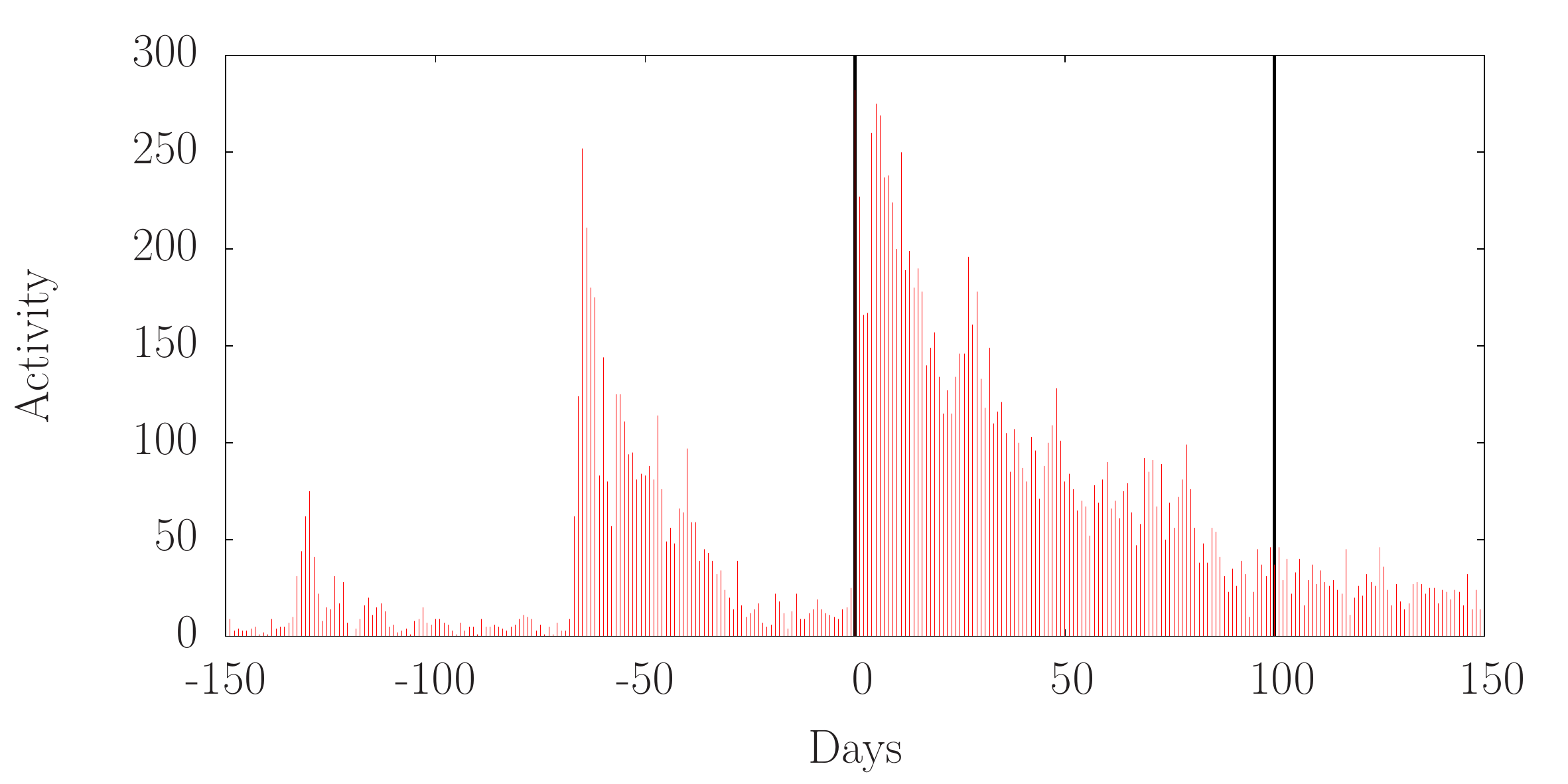}
\caption{\label{fig:2} Number of earthquakes per day of any magnitude before and after the Landers mainshock in the YHS catalog in the area selected for our study. Black lines show the temporal window chosen in this work.}
\end{figure}

\begin{figure}[htpb]
  \centering
  \includegraphics[width=\textwidth]{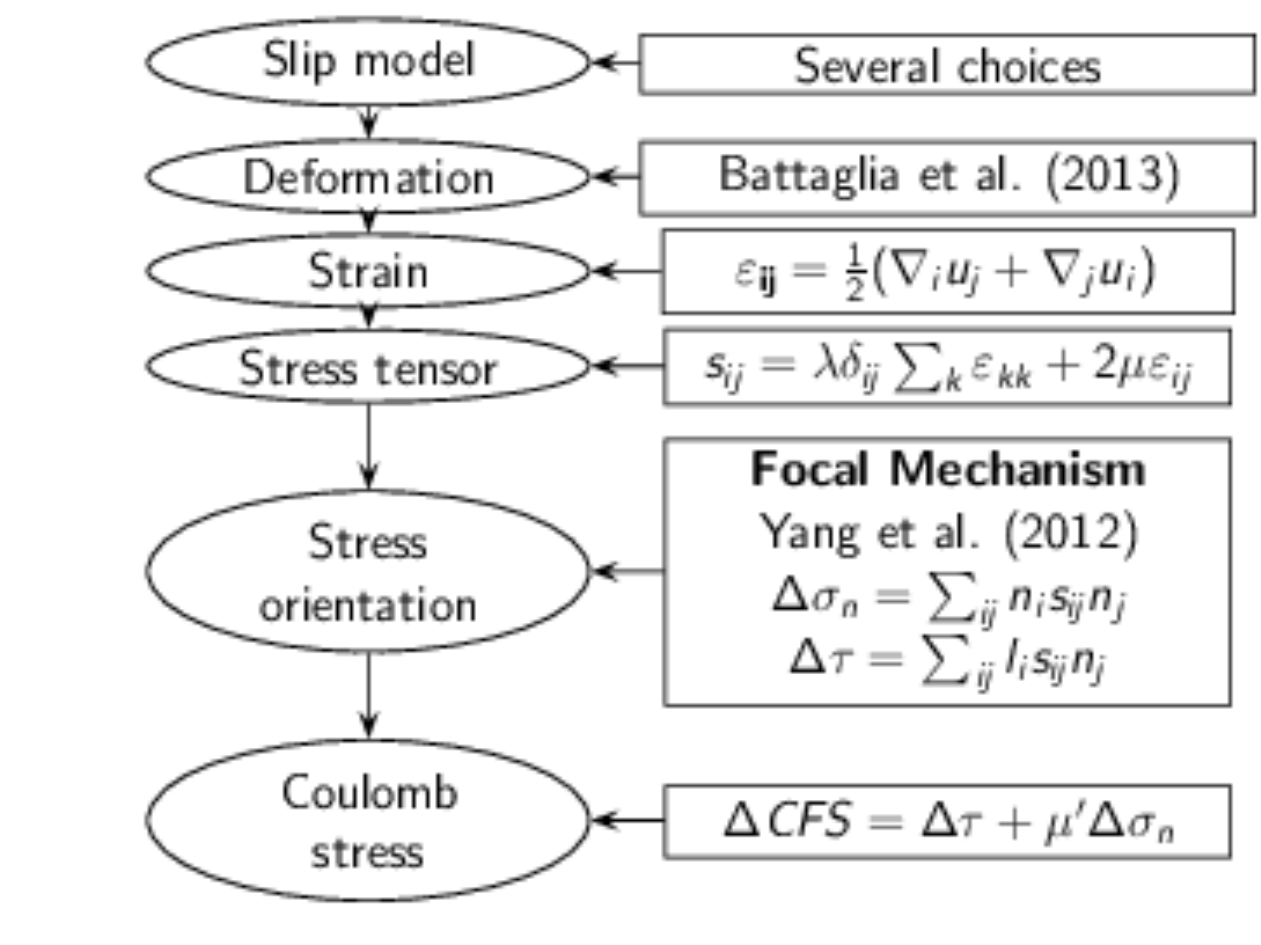}
  \caption{Flowchart summarizing the procedure to obtain the Coulomb stress on each
aftershock fault plane from the slip model and the focal-mechanism catalog.
} 
\label{fig:flujo} 
\end{figure}

\begin{figure}[htpb]
\includegraphics[scale=0.75]{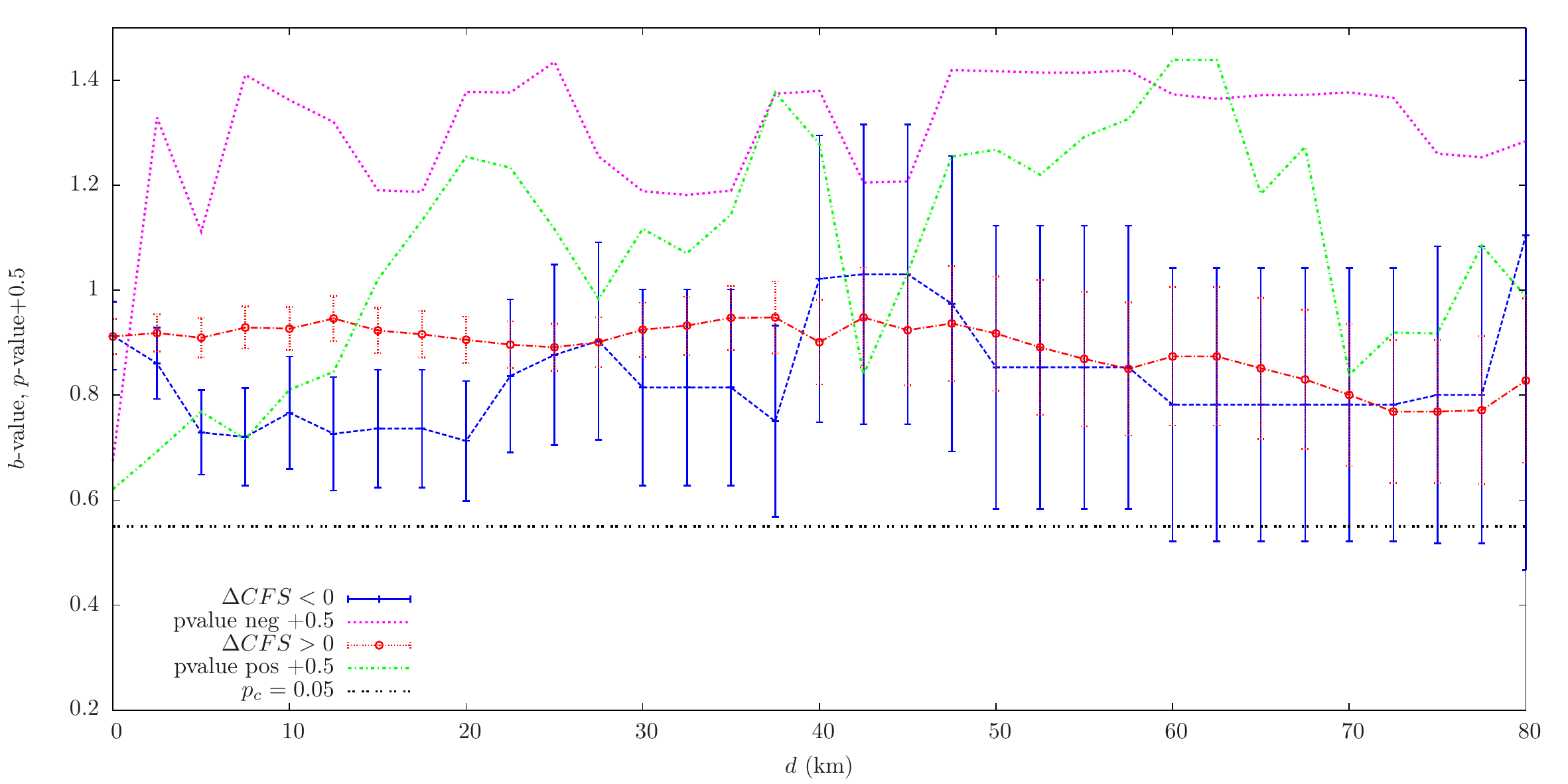}
\caption{\label{fig:1} Dependence of the exponents $b_{>}$ and $b_{<}$ (blue and red respectively) on the distance of the events to the fault $d$ for the wald slip model, with $m_{min}=3$ and $\mu'=0.4$. Green and purple dashed lines correspond to the $p$-values of the goodness-of-fit test shifted $0.5$ for convenience. Horizontal black dashed line corresponds to the threshold $p$-value $p_{c}=0.05$ shifted $0.5$.}
\end{figure}

\begin{table}[htbp]
\center
\begin{tabular}{|c|r|r|r|r|r|}
\hline
%% & \textbf{$d=5km$} & \multicolumn{1}{l|}{} & \multicolumn{1}{l|}{} & \multicolumn{1}{l|}{} & \multicolumn{1}{l|}{} & \multicolumn{1}{l|}{} \\ \hline
  \textbf{} & \multicolumn{1}{l|}{\textbf{$N_{tot}$}} & \multicolumn{1}{l|}{\textbf{${N}$}} & \multicolumn{1}{l|}{\textbf{$b-$value}} & \multicolumn{1}{l|}{\textbf{$\sigma$}} & \multicolumn{1}{l|}{\textbf{$p_{fit}$}} \\ \hline \hline
  Overall & 6730 & 662 & $b_{all}=0.882$ & 0.034 & $0.233 \pm 0.004$ \\ \hline \hline
%%%%%%%%%%%%%%%
\textbf{$\mu'=0.1$,}  \textbf{$\Delta CFS>0$} & 5554 & 564 & $b_>=0.906$ & 0.038 & $0.385 \pm 0.005$ \\ %\hline
\phantom{\textbf{$\mu'=0.1$,}}  \textbf{$\Delta CFS<0$} & 1176 & 98 & $b_<=0.766$ & 0.077  & $0.141 \pm 0.003$ \\ \hline
\textbf{$\mu'=0.2$,}  \textbf{$\Delta CFS>0$} & 5632 & 573 & $b_>=0.900$ & 0.038 & $0.255 \pm 0.004$ \\ %\hline
\phantom{\textbf{$\mu'=0.2$,}}  \textbf{$\Delta CFS<0$} & 1098 & 89 & $b_<=0.739$ & 0.078 & $0.321 \pm 0.005$ \\ \hline
\textbf{$\mu'=0.4$,}  \textbf{$\Delta CFS>0$} & 5678 & 580 & $b_>=0.909$ & 0.038 & $0.267 \pm 0.004$ \\ %\hline
\phantom{\textbf{$\mu'=0.4$,}}  \textbf{$\Delta CFS<0$} & 1052 & 82 & $b_<=0.729$ & 0.081 & $0.617 \pm 0.005$ \\ \hline
\textbf{$\mu'=0.6$,}  \textbf{$\Delta CFS>0$} & 5670 & 582 & $b_>=0.909$ & 0.038 & $0.264 \pm 0.004$ \\ %\hline
\phantom{\textbf{$\mu'=0.6$,}} \textbf{$\Delta CFS<0$} & 1060 & 80 & $b_<=0.730$ & 0.082 & $0.416 \pm 0.005$ \\ \hline
\textbf{$\mu'=0.8$,}  \textbf{$\Delta CFS>0$} & 5603 & 583 & $b_>=0.906$ & 0.038 & $0.263 \pm 0.004$ \\ %\hline
\phantom{\textbf{$\mu'=0.8$,}} \textbf{$\Delta CFS<0$} & 1127 & 79 & $b_<=0.739$ & 0.083 & $0.286 \pm 0.005$ \\ \hline
\hline
 %%%%%%%%%%%%%
% & all & 6730 & 662 & 0.882 & 0.034 & $0.233 \pm 0.004$ \\ \hline
%%%%%%%%%%%%%
 % \textbf{$\delta =60$} 
%& \textbf{$\lambda = 90 \pm \delta$} 
% $\phantom{-1}30^\circ \le \rho \le 150^\circ$ & 1170 & 47 & 0.963 & 0.14 & $0.546 \pm 0.005$ \\ %\hline
%% & \textbf{$\lambda = -90 \pm \delta$} 
% $<150^\circ \le \rho \le -30^\circ$ & 3069 & 93 & 0.964 & 0.10 & $0.666 \pm 0.005$ \\ \hline
\end{tabular}
\caption{
Results of fitting the 
Gutenberg-Richter law to the Landers aftershocks, separating positive and negative Coulomb-stress increases as arising from the wald slip model,
for different values of the effective friction coefficient $\mu'$
and $m_{min}=3$.
The overall case (with $\Delta CFS$ taking any sign)
is also included and labelled as ``all''.
Aftershocks correspond to the first 100 days after the Landers mainshock.
Distance of aftershocks to the Landers rupture is restricted to be between 5 and 150 km.
%The value of $\mu'$ has little effect of the $b-$value.
The $p-$value of the goodness-of-fit test is computed with $10^4$ simulations
and is denoted by $p_{fit}$. 
Its uncertainty corresponds to one standard deviation. We conclude that the value of $\mu '$ has little influence on $b_>$ and $b_<$.
}
\label{tab:mlefitting2}
\end{table}

\section*{Acknowledgements}

We are grateful to \'Alvaro Gonz\'alez for some comments on the manuscript, to Paolo Gasperini, Francesco Mulargia, Fabio Romanelli and Roberto Sabadini for the feedback provided during the International Workshop on Seismic Source Physics, Sardinia, and to two anonymous referees who detected
some problems in a previous version of the manuscript.
The research leading to these results has received founding
from ``La Caixa'' Foundation. V. N. acknowledges financial
support from the Spanish Ministry of Economy and Competitiveness
(MINECO, Spain), through the ``Mar\'{\i}a de Maeztu'' Programme
for Units of Excellence in R \& D (Grant No. MDM-2014-0445).
We also acknowledge financial support from MINECO and MICIU under 
Grants No. FIS2015-71851-P, FIS-PGC2018-099629-B-I00, and 
``Proyecto Redes de Excelencia'' Grant No. MAT2015-69777-REDT.
A. J. appreciates the hospitality of the 
Centre de Recerca Matem\`{a}tica.

\section*{Author contributions statement}

V. N.-P. and A. J. performed the computations, A. J and A. C. wrote the manuscript. All authors discussed the results and reviewed the manuscript. 

\section*{Competing interests}

The authors declare no competing financial and non-financial interests.

\end{document}